# On the energy of electric field in hydrogen atom

## Yuri Kornyushin

Maître Jean Brunschvig Research Unit, Chalet Shalva, Randogne, CH-3975

mailto: jacqie@bluewin.ch

**Summary.** It is shown that hydrogen atom is a unique object in physics having negative energy of electric field, which is present in the atom. This refers also to some hydrogen-type atoms: hydrogen anti-atom, atom composed of proton and antiproton, and positronium.

## 1. Energy of electric field in classical physics

Energy of electric field in classical physics is as follows [1]

$$W = (1/8\pi)\int \mathbf{E}^2(\mathbf{r})dv, \qquad (1)$$

where $\mathbf{E}(\mathbf{r})$ is the vector of the electric field in every point $\mathbf{r}$; integral in taken over the whole space.

When $\mathbf{E}(\mathbf{r})$ is a real-valued quantity, $\mathbf{E}^2(\mathbf{r})$ is positive together with the energy (1).

## 2. Self-action

Let us consider Schrödinger equation, describing some particle [2]:

$$-(\hbar^2/2m)\Delta\psi + U(\mathbf{r})\psi(\mathbf{r}) = E\psi(\mathbf{r}), \qquad (2)$$

where $\psi(\mathbf{r})$ is so called $\psi$-function, $\hbar$ is Planck constant, divided by $2\pi$, $m$ is a mass of a particle, Laplace operator $\Delta = \partial^2/\partial x^2 + \partial^2/\partial y^2 + \partial^2/\partial z^2$, where $x$, $y$ and $z$ are Descartes coordinates, and $E$ is the energy of a stationary state.

Potential energy of a particle in every point $\mathbf{r}$, $U(\mathbf{r})$, is an external potential energy. It does not include interaction of the particle with the field, produced by this same particle. It means that self-action is never taken into account in quantum mechanics.

Averaging quantum-mechanically Schrödinger equation, we have:

$$E = \langle T \rangle + \langle U(\mathbf{r}) \rangle, \qquad (3)$$

where $\langle T \rangle$ is quantum-mechanically averaged kinetic energy operator $T$, and $\langle U(\mathbf{r}) \rangle$ is quantum-mechanically averaged potential energy of a particle in external (with respect to this particle) field, $U(\mathbf{r})$.

The energy of the particle (3) does not include the potential energy of this particle itself. Let us consider a charged particle, which produces electric field. The energy of this field is not included in the energy of a particle (3). This is so because the energy of the electric field produced by the particle is not present in Schrödinger equation. It should be noted that this refers only to the particles charged with elementary charge $\pm e$, such as electron, proton, positron, and antiproton. When regarded particle consists of, for example, two protons, the field produced by each proton is an external one with respect to the other proton and the potential energy of the interaction of the two protons should be introduced in Schrödinger equation.

**3. Energy of electric field in a hydrogen atom**

Electric field in a hydrogen atom is a sum of electric field produced by the charge of electron and electric field produced by the charge of proton, **E** = **E**$_e$ + **E**$_p$. The energy of this field according to (1) is the energy of the field produced by electron plus the energy of the field produced by proton plus interaction energy. As was mentioned before, in quantum mechanics the energy of any electron and any proton are equal to zero. So the energy of the electric field in a hydrogen atom in quantum mechanics consists of the interaction energy only. This interaction energy is negative, because the two particles have charges of the opposite signs. The value of the electrostatic energy of the electric field in a hydrogen atom in the ground state is $U_0 = -(me^4/\hbar^2)$ [3]. Anyway, the energy of the electric field is negative in any stationary state of a hydrogen atom.

**4. Discussion**

Hydrogen atom is a unique object in physics, having negative energy of electric field, which is present in the atom. This refers also to some hydrogen-type atoms: hydrogen anti-atom, atom composed of proton and antiproton, and positronium. In a positronium (a hydrogen type atom, which consists of a positron and electron) the masses of both particles are equal. The charges of the particles are of the same value, but of the opposite signs. That is why the wave functions of both particles are identical. From this follows that in a positronium in a ground state there is no local electric charge, no electric field, but there is negative electrostatic energy.